\documentclass[10pt, conference, compsocconf]{IEEEtran}
\ifCLASSOPTIONcompsoc
  \usepackage[nocompress]{cite}
\else
  \usepackage{cite}
\fi
\ifCLASSINFOpdf
  \usepackage[pdftex]{graphicx}
\else
\fi
\usepackage{amsmath}
\ifCLASSOPTIONcompsoc
  \usepackage[caption=false,font=footnotesize,labelfont=sf,textfont=sf]{subfig}
\else
  \usepackage[caption=false,font=footnotesize]{subfig}
\fi
\usepackage{amssymb}
\usepackage{booktabs}

\newcommand\blfootnote[1]{%
  \begingroup
  \renewcommand\thefootnote{}\footnote{#1}%
  \addtocounter{footnote}{-1}%
  \endgroup
}

\begin{document}
%
\title{Adversarial Attacks on Time-Series Intrusion Detection for Industrial Control Systems}



%
\author{\IEEEauthorblockN{Giulio Zizzo\IEEEauthorrefmark{1}\IEEEauthorrefmark{2}, Chris Hankin\IEEEauthorrefmark{1}\IEEEauthorrefmark{2}, Sergio Maffeis\IEEEauthorrefmark{2} and Kevin Jones\IEEEauthorrefmark{3}}
	\IEEEauthorblockA{\IEEEauthorrefmark{1}Institute for Security Science and Technology (ISST), Imperial College London}
	\IEEEauthorblockA{\IEEEauthorrefmark{2}Department of Computing, Imperial College London}
	\IEEEauthorblockA{\IEEEauthorrefmark{3}Airbus}
	\IEEEauthorblockA{g.zizzo17@imperial.ac.uk\qquad c.hankin@imperial.ac.uk\qquad sergio.maffeis@imperial.ac.uk\qquad kevin.jones@airbus.com}
	}


\maketitle

\begin{abstract}
Neural networks are increasingly used for intrusion detection on industrial control systems (ICS). With neural networks being vulnerable to adversarial examples, attackers who wish to cause damage to an ICS can attempt to hide their attacks from detection by using adversarial example techniques. In this work we address the domain specific challenges of constructing such attacks against autoregressive based intrusion detection systems (IDS) in an ICS setting.

We model an attacker that can compromise a subset of sensors in a ICS which has a LSTM based IDS. The attacker manipulates the data sent to the IDS, and seeks to hide the presence of real cyber-physical attacks occurring in the ICS.

We evaluate our adversarial attack methodology on the Secure Water Treatment system when examining solely continuous data, and on data containing a mixture of discrete and continuous variables. In the continuous data domain our attack successfully hides the cyber-physical attacks requiring 2.87 out of 12 monitored sensors to be compromised on average. With both discrete and continuous data our attack required, on average, 3.74 out of 26 monitored sensors to be compromised.

\end{abstract}


%
\IEEEpeerreviewmaketitle

\section{Introduction}\blfootnote{Accepted at the 2020 IEEE 19th International Conference on Trust, Security and Privacy in Computing and Communications (TrustCom). DOI: 10.1109/TrustCom50675.2020.00121}
Deep learning systems are known to be vulnerable to adversarial attacks. By applying small changes to an input, an attacker can cause a machine learning system to misclassify with a high degree of success. There has been much work on both developing more powerful attacks \cite{carlini2017towards} as well as defences \cite{madry2017towards}. However, the majority of adversarial machine learning research is focused on the image domain, with consideration of the challenges that arise within other fields needed \cite{carlini2018audio, pierazzi2020intriguing}.

The phenomenon of adversarial examples becomes particularly pertinent when aiming to defend machine learning systems operating as security solutions. Machine learning systems frequently outperform other methods in detecting cyber attacks\,\cite{kim2016lstm, KevinandRhode2017early, grosse2016adversarial, javaid2016deep}. Despite this advantage, vulnerability to adversarial examples means that adaptive attackers can pose an immediate risk. 

We consider the problem of adversarial examples targeting intrusion detection for industrial control systems (ICS), as it is a domain in which machine learning systems are being both researched \cite{kim2016lstm, inoue2017anomaly} and offered by vendors as security solutions. In particular, we explore ICS adversarial vulnerabilities under $L_0$ constraints. This involves the examination of an attacker which can compromise sensors and actuators in an ICS and substitute their readings, which are processed by an intrusion detection system (IDS), with attacker controlled data. It can be trivial to compromise an IDS if the attacker can control a large portion of the data. The challenge is to achieve evasion with a minimal number of compromised sensors and actuators.

The contributions of this paper are as follows: 

\begin{itemize}
    \item We demonstrate the vulnerability of time-series based intrusion detection operating on a ICS to adversarial examples, and demonstrate how to hide a range of real cyber-physical attacks. 
    \item We examine challenges in attacking an autoregressive IDS due to attacker perturbations propagating to the optimisation target. We analyse this via loss landscape visualisations of adversarial examples.
\end{itemize}

\section{Background}

\subsection{Adversarial Examples}\blfootnote{© 2020 IEEE.  Personal use of this material is permitted.  Permission from IEEE must be obtained for all other uses, in any current or future media, including reprinting/republishing this material for advertising or promotional purposes, creating new collective works, for resale or redistribution to servers or lists, or reuse of any copyrighted component of this work in other works.} An adversarial example is data that has been manipulated to cross the decision boundaries of a machine learning model in order to be mis-classified or, for regression tasks, control the model predictions. The concept of the \textit{action space}~\cite{gilmer2018motivating} of the attacker determines the allowable perturbations. For images this is typically taken as adding imperceptible perturbations to the image (however in patch based attacks the perturbation is often visible~\cite{brown2017adversarial, karmon2018lavan}). Outside the image domain different action spaces are more appropriate, as human perception and semantic meaning have less relevance. In particular are action spaces with \textit{content constraints}. These are constraints that arise from the data, for example in hiding malware the original attack code must still function~\cite{suciu2018exploring}. In the ICS domain we find ourselves closer to the content constrained attack model compared to the indistinguishable perturbation model used for images. Our adversarial perturbations added for ICS are visible, and we can only add the perturbations to particular features as dictated by our attacker model (for example compromising the data leaving a sensor). Furthermore, the underlying data has less obvious semantic meaning to a human. While the content of a picture is immediately obvious, the semantic meaning of ICS data is harder to determine as it will depend on the overall status of the system which is composed of many sensors and actuators.

When conducting an evasion attack, an attacker adds a perturbation to a datapoint $x$ such that a neural network will output different classes for $x$ and the perturbed datapoint $\hat{x}$. There are a range of attack algorithms to craft the adversarial perturbation operating under different bounds. For example, patch attacks, representing a $L_0$ bound, can be used to create adversarial stickers~\cite{brown2017adversarial} and street signs~\cite{eykholt2018robust}. In terms of $L_{\infty}$ bounds the Fast Gradient Sign Method (FGSM)\cite{szegedy2013intriguing}, can quickly generate adversarial samples by perturbing each pixel in an image. On the other end of the spectrum, in terms of attacker strength, the Carlini Wagner attack \cite{carlini2017towards} optimises for data misclassification and simultaneously keeping the perturbation as small as possible.

Defences for adversarial examples are an active research area using ideas from robust training~\cite{madry2017towards}, uncertainty~\cite{feinman2017detecting}, intermediate layer activations~\cite{papernot2018deep,zizzo2019deep}, or removing adversarial perturbations~\cite{xu2017feature}. However, there is still no silver bullet to defending against all adversarial examples.

\subsection{Related Work for Intrusion Detection}

Machine learning can be used to detect anomalies in industrial control systems. Such detection systems are often autoregressive. Based on data at time steps $x_{0}, \dots , x_{t}$ a prediction $y_{t}$ is made for $x_{t+1}$. This prediction is compared to what is actually observed and the difference forms a residual, $r_{t+1}$. A detection function $F_d$ is then employed to determine if an attack is occurring. $F_d$ represents any operations which are computed on the residuals to generate an alert; examples include averaging, smoothing, and thresholding. The exact machine learning model and detection function vary across different works.  

Long short-term memory (LSTM) networks were investigated in \cite{kravchik2018detecting, inoue2017anomaly, goh2017anomaly} for detecting cyber-physical attacks in the Secure Water Treatment (SWaT) system with the best LSTM achieving a $F_1$ score of 0.802. A different approach was taken in \cite{feng2017multi} which investigated the Gas Pipeline dataset \cite{morris2015industrial} and combined a filtering step followed by an LSTM prediction stage. Many other models have been analysed. Convolution neural networks~\cite{kravchik2018detecting} have reached high $F_1$ scores and autoencoders have been investigated in~\cite{kravchik2019efficient, taormina2018deep} as well as neural architecture search in \cite{shalyga2018anomaly}. Alternatively, ensembles of random trees are used in \cite{hassan2020increasing}, or a range of classification methods including SVMs, neural networks, and tree based methods are compared in \cite{hassan2020adaptive}.

Recent work in \cite{erba2019real} generated adversarial attacks for ICS when attacking an autoencoder IDS. Their attacker substitutes the original data for readings within normal sensor range. The perturbation applied in this way could be extremely large, as every sensor reading could be replaced from an arbitrary initial value, to a value that is within normal sensor range. Alternatively, generative methods can be used to create adversarial attacks such as in \cite{feng2017deep} where generative adversarial networks (GANs) were used to create adversarial data. Additional work in \cite{kravchik2019efficient} investigated adversarial attacks targeting an autoencoder IDS. Unlike \cite{erba2019real}, the work in \cite{kravchik2019efficient} modelled their attacker as not having control of the communications to the IDS independently of the programmable logic controller (PLC). Therefore, the adversarial data had to fool the IDS and fulfill the original cyber-physical attack. With this objective, even with white box knowledge of the IDS and perfect knowledge of future system states, an effective adversarial attack was not found.

\section{Attacker Model}

We first introduce our adaptive adversarial attacker model in terms of the goals the attacker wishes to achieve, their capabilities in manipulating the system, and finally the overall knowledge of the system they possess. 

\subsection{Attacker Goals}

The attacker in our situation wishes to conduct a cyber-physical attack on an ICS. However, IDS solutions can quickly detect these attacks, and prevent damage. Our attacker aims to conduct the cyber-physical attacks while remaining hidden from an IDS for the attack's entire duration. 

\subsection{Attacker Capabilities}

To achieve their aim we assume the attacker is able to control the data flow between $k$ sensors or actuators to an IDS. By sending tampered data to an IDS the attacker aims to hide the cyber-physical attacks. Effectively, the attacker is operating over a restricted $L_0$ constraint. An $L_0$ constraint specifies how many features a attacker can modify. Normally the attacker can modify any set of features they choose, as long as it is smaller than the specified $L_0$  bound. However, in addition to this $L_0$ constraint our attacker is further restricted as they can only modify features on a sensor by sensor basis. In other words, the attacker cannot split their $L_0$ perturbation budget across multiple sensors and actuators without also requiring that those sensors/actuators become fully compromised. 

An example in practice for the ICS we will later examine involves an attack in which water level as measured by a sensor rapidly changes value. If the attacker can now control the data leaving that sensor to an IDS, how can the data be altered in order to hide their attack? Simply reporting a constant fixed water level of the original measurement still triggers an alarm as the senor values are not consistent with the rest of the system dynamics. It becomes even less intuitive if the adversary cannot send false sensor data on certain features. As an example, if the attacker turns on a pump to cause a cyber-physical attack, and this pump state cannot be manipulated further, which of the sensors that triggers alerts should the attacker compromise, and what is the perturbation that should be added?

Finally, in a typical ICS, sensors and actuators communicate via a series of PLCs to the IDS. We examine the direct compromise of individual sensor readings, instead of a whole PLC, as this offers finer granularity to assess the sensitivity of adversarial attacks. For example, we show that a cyber-physical attack can be hidden by adversarial optimisation of a single flow sensor. On a PLC basis this would mean compromising the whole of a particular PLC, which controls many different sensors and actuators, missing the information that flow sensor is the weakest link.

\subsection{Attacker Knowledge}
\label{subsec:Attacker_Knowledge}
We model the attacker as having white box knowledge of an IDS and the physical system dynamics. Knowledge of the physical system dynamics has a large influence on the attacker strength. If the attacker does not know how the system evolves in time, then all the attacker would be able to do is greedily optimise data as it arrives to minimise the current residual. After the attacker has sent the manipulated data to an IDS they will be unable to retroactively make changes to aid them in keeping future attack data hidden. 

If the attacker has a model of the system then they can predict how it will evolve over the course of an attack. With this capability the attacker optimizes the data with the aim of stealth across all time steps. We represent this capability by assuming the attacker's model of the physical system yields its ground truth, but cannot generate the appropriate noise that will occur when a sensor takes a measurement. 

To simulate the correct level of attacker knowledge when only having access to datasets with existing sensor noise a degree of approximation is required. We therefore take the values in a ICS dataset as the ground truth of the physical system as given by the attacker's model. This dataset, $D_a$, is the knowledge the attacker has of how the system evolves. Then, we create a copy of the dataset and add the appropriate level of Gaussian sensor noise to it. This set of data, $D_d$, is what the defender will see. Thus $D_a$ and $D_d$ differ by the level of Gaussian noise that the attacker cannot predict.

Having an accurate model of the physical system is a significant requirement, however it is an accepted assumption in this area. 
Related work provided the attacker with knowledge of all of the data, free of any noise~\cite{kravchik2019efficient}, or considered defences which evaluate datapoints independently, and enable attackers to compromise a target without knowledge of the physical system~\cite{erba2019real}.
In comparison our model is more challenging for the attacker.

\section{Attack Algorithms}

We introduce two algorithms that can be used to evade an autoregressive based IDS. The first is an optimisation based strategy which perturbs the attacker controlled features with the aim of stealth across all time steps, and is not limited to autoregressive systems. The second is specific to autoregressive based systems but has the strong advantage of requiring little system knowledge aside from the IDS model itself. 

\subsection{$L_{0}$ Optimisation Attack}
\label{L_0_Opt}
In the optimisation attack, the attacker sends adversarial data on the compromised features to an IDS to hide cyber-physical attacks across all monitored features. The IDS generates residuals $r_{1} \dots r_{N}$ which are passed though a detection function $F_d$ which outputs zero if no attack is detected or $\eta$, the magnitude by which an alert is generated. 
The attacker's goal is thus to minimise the detection loss $L_D$:
\begin{equation}
    L_D = F_d(r_{1}\dots r_{N}).
\end{equation}

\begin{figure*}
    \subfloat[]{\includegraphics[trim = 0mm 0mm 110mm 0mm, clip, scale=0.4]{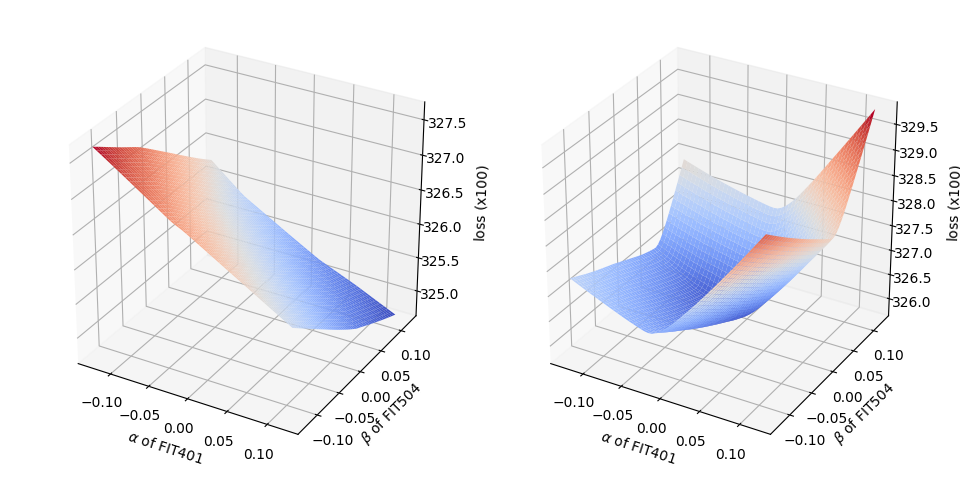}}
    \subfloat[]{\includegraphics[trim = 130mm 0mm 0mm 0mm, clip, scale=0.4]{Visualisations/Attack_33.png}}\hspace{5mm}
    \captionsetup[subfloat]{width=-0.4cm}
    \subfloat[]{\includegraphics[trim = 0mm 0mm 100mm 0mm, clip, scale=0.35]{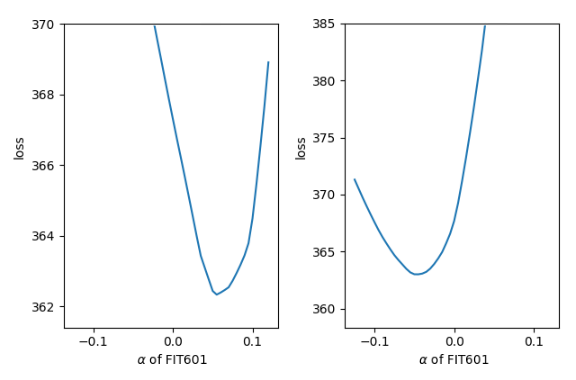}}
    \subfloat[]{\includegraphics[trim = 100mm 0mm 0mm 0mm, clip, scale=0.35]{Visualisations/Attack_2.png}}
    
    \caption{Graphs illustrating the effects of the change in optimisation target between $\hat{\mathbf{T}}$ and $\hat{\mathbf{T}}^*$. The horizontal axes show the magnitude of perturbation applied on a specific sensor. On the leftmost plot we have the loss surface when applying a range of gradient based perturbations on two different sensors FIT401 and FIT504, which are both flow level sensors. The loss is evaluated on the original target vector, $\hat{\mathbf{T}}$, that the gradient is computed on. Even in the largest perturbation applied the loss decreases by the maximum amount. Then, in (b), we generate the new target vector $\hat{\mathbf{T}}^*$ and compute the loss with  $\hat{\mathbf{x}}^*$ as the input. Rather than the loss decreasing, the loss increases except for the smallest gradient update steps. On the rightmost pair of plots we see a ``worst case" scenario when changing between $\hat{\mathbf{T}}$ and $\hat{\mathbf{T}}^*$. With (c) we have the losses when applying perturbations with $\alpha$ between -0.1 to 0.1 on the flow level sensor FIT601. By applying perturbations up to $\alpha \sim 0.05$ the loss decreases. However, if we apply the perturbation and re-compute the loss with respect to $\hat{\mathbf{T}}^*$ as shown in (d) the perturbations as given by the gradient with positive $\alpha$ on $\hat{\mathbf{T}}$ always gives a higher loss. The best choice is to go in the \textit{opposite} direction to what is expected from the gradient in order to reduce the loss when using $\hat{\mathbf{T}}^*$.}
\label{fig:Attack_33_Two_Comp}

\end{figure*}

Although the attacker can alter the data on the compromised features they do not have complete freedom as the compromised features are themselves monitored. The adversarial datapoint $\hat{x}_{t}$ will cause a autoregressive IDS to make a prediction $y_{t}$ which must be close to the adversarial datapoint $\hat{x}_{t+1}$. However, $\hat{x}_{t+1}$ is changing every iteration step as it is itself being updated. To express this more precisely, a sequence on the optimisation step $i$,
\begin{equation}
    \hat{\mathbf{x}} = \hat{x}^i_{t_0 \dots t_N}
\end{equation}

will be optimised to reduce the loss between IDS predictions $y^i_{t_0 \dots t_N}$ and the target data sequence

\begin{equation}
    \hat{\mathbf{T}} = \hat{x}^i_{t_{1} \dots t_{N+1}}.
\end{equation}

On the next iteration step the adversarial datapoint is updated with the adversarial perturbations, $\delta^i_{t_0 \dots t_N}$,  to

\begin{equation}
    \hat{\mathbf{x}}^* = \hat{x}^i_{t_0 \dots t_N} + \delta^i_{t_0 \dots t_N}
\end{equation}

while the optimisation target is 

\begin{equation}
    \hat{\mathbf{T}}^* = \hat{x}^i_{t_{1} \dots t_{N+1}} + \delta^i_{t_{1} \dots t_{N+1}}.
\end{equation}

So, although we take an optimisation step towards minimising the loss with respect to $\hat{\mathbf{T}}$, our new loss is computed on $\hat{\mathbf{T}}^*$ and the steps $\delta^i$ may not be the correct ones to take with respect to these new targets. This creates a situation that is not usually present in static data like images, as in our case the optimisation target for the attacker changes from iteration step to iteration step. 

This presents a challenging optimisation problem. To explore its effects, and develop a strategy for solving it we can look at loss surfaces for our adversarial examples. Examining the behaviour of the loss around high dimensional objects such as adversarial examples involves a projection onto lower dimensions. The simplest way to achieve this is via interpolation of the loss $L$ between the adversarial example $\hat{x}$ and $\hat{x} + \epsilon$ where $\epsilon$ is a direction vector. We can explore 2D surface plots by introducing a second vector $\nu$ and weighting parameters $\alpha$ and $\beta$ and plot:

\begin{equation}
     \label{equ:loss_ladscape}
    f(\alpha, \beta) = L(\hat{x} + \alpha\epsilon + \beta\nu).
\end{equation}

When dealing with the loss landscape around adversarial examples we select directions based on the perturbation direction that the attacker will add, i.e the sign of the gradient with respect to the loss. Thus, we can set $\epsilon$ and $\nu$ to be vectors given by the sign of the gradient of the compromised sensors with respect to the loss.

We plot the loss landscapes in Figure \ref{fig:Attack_33_Two_Comp} for cyber-physical attacks in the SWaT dataset that we later examine\footnote{The nomenclature of SWaT components is reported in Appendix \ref{sec:Naming_Conventions}.}. In Figure \ref{fig:Attack_33_Two_Comp}a we see that if we conduct gradient descent the loss when evaluated on the original targets $\hat{\mathbf{T}}$, even for large $\alpha$ and $\beta$, decreases. When the target is updated to $\hat{\mathbf{T}}^*$, shown in Figure \ref{fig:Attack_33_Two_Comp}b, the targets have changed enough such that our loss is higher than when we started for a large range of values of $\alpha$ and $\beta$. We see that the optimal action with respect to $\hat{\mathbf{T}}$ places us in a higher state of loss compared to our starting location ($\alpha = 0 $ and $\beta = 0$) when updating the target to $\hat{\mathbf{T}}^*$. It is important to note that it is not a simple case of the learning rate being too high. The loss with respect to the original target, $\hat{\mathbf{T}}$, consistently reduces even when using large updates. Rather, the change between $\hat{\mathbf{T}}$ and $\hat{\mathbf{T}}^*$ alters the objective such that we can easily end up in a poor location with respect to the new targets.

This can have a very detrimental effect on optimiser performance. In the case of Figures \ref{fig:Attack_33_Two_Comp}c and \ref{fig:Attack_33_Two_Comp}d applying perturbations based on the correct gradient direction results in the loss rising when new predictions are compared to $\hat{\mathbf{T}}^*$. In fact, to reduce the loss in that case, the best action is to follow the gradient in its \textit{opposite} direction to make the loss with respect to $\hat{\mathbf{T}}^*$ lower than the loss we began with on $\hat{\mathbf{T}}$.

The BFGS optimiser proved the most robust and quickest amongst the optimisers we evaluated. Figure \ref{fig:Attack_33_Two_Comp} offers insights as to why. The BFGS method, after obtaining a search direction, performs a line search to find the most appropriate step-size thus avoiding potential situations like that in Figure~\ref{fig:Attack_33_Two_Comp}. For this domain therefore, an optimisation strategy \textit{must} incorporate an adaptive search for the step size, in contrast to many other domains with adversarial examples for which simpler optimisation strategies are effective.

\subsection{$L_{0}$ Prediction Attack}

This attack exploits the autoregressive nature of an IDS and functions by feeding the IDS's predictions back to itself as though it is real sensor data. Concretely, if the attacker has the neural network model, then at time $t$ they can compute the IDS's predictions for the next time step $y_{t}$. Then, at $t+1$, on the features the attacker controls, rather than sending the real data, $x_{t+1}$, they send $y_{t}$ to the IDS- i.e at every time step they perfectly match the data they send to the IDS with its own prediction for the system state. 

This attack method does not make active attempts to reduce detection on additional monitored features, and so hiding attacks which have been detected on more features than the attacker has compromised is not guaranteed to succeed. However, this method does have advantages for the attacker as the knowledge they need of the target system is lower. The detection function, $F_d$, and a model of the physical system dynamics are not needed to run the attack. 

This method can be combined with the $L_{0}$ optimization attack by first running the prediction attack which provides an initialisation closer to the final goal. This often led to better results in comparison to starting the optimisation attack from the original data.

\section{Adversarial Attack Implementation}

\subsection{Attack Strategy}
\label{sec:Attack_Strategy}

To generate adversarial examples for cyber-physical attacks we begin optimising $T$ time-steps before the cyber-physical attack begins to aid in hiding the initial portion of the attack. The value of $T$ will be context dependent on the particular ICS being examined, in fast moving systems $T$ could be a short time frame while in slow processes $T$ may need to represent a long time period. We examine the SWaT system~\cite{goh2016dataset}, and select $T$ to be 400, 100 or 25 seconds prior to an attack starting. The largest window is chosen which did not begin during a previous attack.

We examine our attacks in two cases, the first when an IDS only monitors continuous data, and the second where an IDS monitors all features, both continuous and discrete. The motivation for this distinction is that both scenarios could be realistically encountered by an attacker. Some ICS or sensor networks can be composed of purely continuous data streams, whilst other setups would have both continuous sensor measurements and discrete actuator states. 

When an IDS monitors only continuous sensors we explore three strategies for each cyber-physical attack. These can be ranked from the most preferable to least: 

\begin{enumerate}
    \item Apply $L_0$ Prediction attack: This attack requires only the IDS model to function without requiring a physical system model or the detection function employed.
    \item Apply $L_0$ Prediction followed by $L_0$ Optimisation: Knowledge of the system model and the detection function is needed in addition to the IDS model, but is frequently faster than purely $L_0$ Optimisation.  
    \item Apply $L_0$ Optimisation attack: Requires a high level of knowledge and is frequently the most computationally expensive. 
\end{enumerate}

The attack strategy which requires the smallest level of system compromise is then selected. If two or more strategies require equal level of compromise the one with the better ranking as defined above is used.

When both continuous and discrete data-types are monitored we always initialise compromised features with the $L_0$ Prediction attack which functions on both data-types effectively. Then, we fix the discrete data values and apply the $L_0$ Optimisation attack on only the continuous compromised features. We experimented with different strategies for simultaneously optimising the continuous and discrete features, for example passing the discrete features through a sigmoid function to give gradients for optimisation, however they did not provide benefits over our final strategy.  

For the $L_0$ constrained attacker we need to determine which features the attacker should compromise. We select the first feature to optimise based on which one had the highest detection loss computed on the dataset $D_a$ which the attacker controls. Note, from Section \ref{subsec:Attacker_Knowledge} the dataset $D_a$ differs from the dataset which the defender will see $D_d$ due to the addition of extra sensor noise. 

We then run our selected attack strategy and if the resulting detection loss is greater than zero we increase the level of compromise by including the feature with the highest level of detection loss post-optimisation. This continues iteratively until, based on the attacker dataset $D_a$, the attacker has zero detection loss. 

Once zero detection loss is achieved on $D_a$ we check for pruning of the compromised features. As the compromised feature $k_i$ is selected conditioned on the compromised features $k_0 , \,\dots\,,  k_{i-1}$ the effect of more recent compromised features $k_{i+1} , \,\dots\,,  k_{i+n}$ may make the compromise of feature $k_i$ unneeded. Thus, we iterate backwards removing compromised features beginning at $k_{i+n-1}$ and checking that zero detection loss is still achieved. If it is achieved we remove $k_i$ from the features needing compromise.

\subsection{Replay Attack}

We use a replay attack as a simple baseline comparison. In a replay attack the attacker has access to all prior data. The attacker substitutes data on any features they control with data they have recorded. As ICS are frequently periodic the attacker substitutes in data gathered at the same time $n$ days prior. A similar attack strategy is adopted in \cite{erba2019real}. $n$ is the smallest integer for which the substituted data contains no anomalies. The features to compromise are selected in the same manner as our machine learning based attacks. 

\subsection{Effects of Sensor Noise}
\label{sec:sensor_noise}
Once a sequence of adversarial data is computed on the compromised features it replaces the appropriate features in the defender's dataset $D_d$ and ran through the IDS. If the attack achieves zero detection loss then the attack is successful. However, due to unknown sensor noise in uncompromised features the attack may fail. This occurs when the attacker achieves zero detection loss on their dataset $D_a$ but when applying the perturbations onto the defender data $D_d$ detection loss is still present. To account for unknown noise the attacker optimises to a fraction $\tau$ of the detection threshold values. The principle behind this is that if an attack can be optimised to lie below $\tau$ of the detection threshold, then even if there is additional noise, it will not drive the cumulative residuals above the full detection threshold. 

There is a trade-off between making the resulting adversarial sample robust to unknown noise and not needlessly requiring additional compromise. We set $\tau$ to 0.9 when attacking purely continuous features and 0.95 when attacking mixed data types as the latter is a more challenging task. We did not run an exhaustive hyperparameter selection of $\tau$ due to the computational time requirements of generating complete sets of adversarial attack sequences. Optimal selection of such hyperparameters is left as future work. 

\section{Defender Model: SWaT Case Study}

To analyze the vulnerability of a time-series based IDS to the described attack algorithms, we use the SWaT dataset which is gathered on a water treatment ICS. Full details on the dataset can be found~\cite{goh2016dataset} and \cite{SWaT_Website}. There have been several detection systems proposed for the SWaT dataset \cite{goh2017anomaly, kravchik2018detecting, shalyga2018anomaly, li2019mad}. In this work, we train a LSTM \cite{gers2000learning} as the IDS, as it achieves high detection performance, and so represents a useful baseline.

At a high level the SWaT system is a 6 stage scaled down water treatment system. It has numerous sensors measuring physical properties, such as water flow rate in pipes and water level height in tanks. Additionally, chemical monitoring sensors measure a range of characteristics such as water conductivity and pH. Finally, there are numerous actuators which control the water flow rate and chemical dosing.

The dataset was gathered over a period of 11 days. Over that time, 7 days were run under normal system operation and over the course of 4 days a total of 36 different attacks were run on the SWaT testbed. These attacks differed in duration and objective, with some attacks seeking to create underflow/overflow situations in water treatment tanks, while others aimed to burst pipes and halt filtration processes. The data itself is comprised of the sensor and actuator measurements extracted from the raw network traffic conducted over industrial EtherNet/IP and Common
Industrial Protocol (CIP) stack.
\subsection{IDS Model}
\label{subsec:Model}
For the LSTM IDS we divide the data into sliding windows of 100 time steps. At each time step, $t$, the LSTM makes a prediction, $y_{t}$, for the next system state, $x_{t+1}$, based on the past datapoints $x_{1}, \,\dots\,, x_{t}$.

We use an LSTM with four layers, each with 512 hidden units and a dropout rate of 0.5 between layers. A set of dense layers takes the LSTM's output at every time step and produces predictions for every feature. 

At test time we take the difference between the predicted and observed values to form a series of residuals, $r_{1} , \,\dots\,, r_{t}$, for every predicted feature. We assume that the errors are normally distributed and so we compute means, $\mu$, and standard deviations, $\sigma$, of residuals on the validation data. The positive, $R^p_t$, and negative, $R^n_t$, residuals are then computed based on 

\begin{equation}
    R^p_t = max(0, r_t - \mu -\sigma)
\end{equation}
\begin{equation}
    R^n_t = min(0, r_t - \mu +\sigma).
\end{equation}

Finally, we perform two cumulative sums over a sliding window containing 10 timesteps for both $R^p_{1\dots t}$ and $R^n_{1\dots t}$. If the cumulative sum exceeds a threshold on any of the predicted features an anomaly is declared.

\subsection{Data Processing}

\begin{figure}[]
    \centering
    \includegraphics[trim = 0mm 0mm 0mm 0mm, clip, scale=0.35]{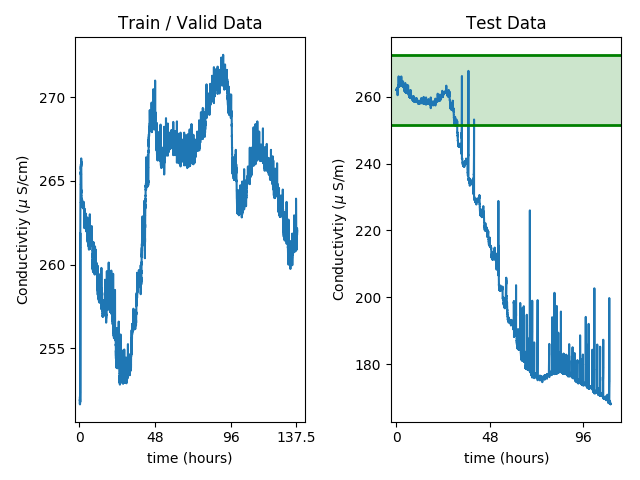}
    \caption{Left: Training data for the AIT201 sensor. Right: Test time data, the green area indicates the range spanned by the training data.}
    \label{fig:AIT201}
\end{figure}

We noticed several features in the attack dataset experienced significant drift in behaviour with respect to the data collected to represent normal system operation. One such example for the sensor AIT201, measuring water conductivity, is shown in Figure \ref{fig:AIT201}. We can see that the test data remains in the range spanned by the training and validation data for a short period. In Figure \ref{fig:AIT201} we removed any data that is associated with a cyber-physical attack, as one may expect attacks to have different statistical properties with respect to normal system operation. Hence, all of the data shown represents normal behaviour. As the test data representing normal system operation has a distribution fundamentally different from the training data it can be flagged as anomalous, despite it not belonging to a cyber-physical attack. It is worth emphasising that this test time data is indeed anomalous with respect to the training data, which is all we have access to a priori. 

Some works use test time statistics to normalise the test data and so sidestep the problem of drift between the two portions of the dataset \cite{li2019mad, inoue2017anomaly} however that introduces data snooping and is not considered best-practice. To address this issue we only used features that relate to physical system properties in our IDS, i.e. flow and water level sensors as well as mechanical pumps and valves. These features did not experience such large deviations compared to many of the sensors relating to chemical measurements.

The authors of~\cite{kravchik2019efficient} also noticed the discrepancy between the test and training datasets for non-anomalous data. In their work they propose the use of a modified Kolmogorov-Smirnov (KS) test to filter out such features. However, in order to remove the features that exhibited large variation between the training and testing datasets it also filters out several physical features which our LSTM is able to effectively model, suggesting that the KS test results in a excessive pruning of features.

\subsection{IDS Evaluation}
\label{sec:IDS_Results}
We train our LSTM and evaluate it on the test set. We feed the data in sequential windows comprised of $T$ seconds to the IDS. If in the most recent window an anomaly is detected we reset the LSTM's internal state, otherwise we use the LSTM in a stateful manner. This allows for fewer cold starts to be experienced by the LSTM over the normal operation of the SWaT system while preventing anomalies from affecting the LSTM's internal state after the attack has ended. We carry out a small grid search of $T = \{25, 50, 100, 150\}$ and 100 gave the best result.     

We tune our detection thresholds by running a grid search over the thresholds for the monitored features. We use $F_1$ score as a performance metric defined as

\begin{equation}
    F_1 = \frac{2TP}{2TP+FP+FN}
\end{equation}
where the true and false positive counts are $TP$ and $FP$ and the number of false negatives are $FN$.

The resulting $F_1$ scores are in Table \ref{tab:F1_Scores} and our method achieves an $F_1$ score of 0.856, making it one of the strongest literature baselines. To enable accurate comparison to previous works we use the original SWaT dataset, rather than $D_d$ which is produced as described in section \ref{subsec:Attacker_Knowledge}.

\begin{table}[]
    \centering
    \caption{Results for different benchmarks. For \cite{kravchik2018detecting} we report the performance of non ensemble methods from the results table}
    \begin{tabular}{c c}
    \toprule
    Method & $F_1$ score  \\
    \hline
    \addlinespace[0.1cm]
     MLP \cite{shalyga2018anomaly} & 0.812 \\
     \addlinespace[0.1cm]
     SVM \cite{inoue2017anomaly} & 0.796\\
     \addlinespace[0.1cm]
     DNN \cite{inoue2017anomaly} & 0.802 \\
     \addlinespace[0.1cm]
     MADGAN \cite{li2019mad} & 0.77 \\
     \addlinespace[0.1cm]
     Various \cite{kravchik2018detecting} & 0.609 - 0.775\\
     \addlinespace[0.1cm]
     Autoencoders \cite{kravchik2019efficient} & 0.873 \\
     \addlinespace[0.1cm]
     \textbf{Ours} & {\textbf{0.856}} \\ 
     \bottomrule
     \addlinespace[0.1cm]
    \end{tabular}
    
    \label{tab:F1_Scores}
\end{table}

\subsection{Stronger Defender Model}
\label{Defender Model}

Within the SWaT dataset several cyber-physical attacks have long term effects on the system which can require significant time to re-stabilize. This data, although anomalous, is not labelled as an attack and so would contribute to the false positive rate. Hence, this results in detection thresholds being artificially raised. We wish to devise an IDS which is free from such negative effects as it presents a target for the attacker which is weakened due to these artefacts which will not be present in all IDS instantiations.

To that end we use the training portion of $D_d$ to train a new IDS model. We now alter our detection thresholds to make the defender stronger against an adversarial attacker.

We extract from the SWaT attack set data that is ``clean" of secondary effects and does not contain any attacks, from which we establish detection thresholds and false positive rates. Clean data sequences are defined by\footnote{An exception to the stated conditions is the data between attacks 23 and 24. The system takes almost all of the duration between those two attacks to return to normal operation and that fragment of data is not used.}:

\begin{enumerate}
    \item Beginning 600 seconds after an attack is labelled as finishing. 
    \item Ending 100 seconds before an attack is labelled as starting, as some attack effects begin a few seconds before an attack is labelled as starting. 
    \item The data sequence is at least 1500 seconds long. 
\end{enumerate}

To determine our attack detection capability we then extract sequences of data which contain the cyber-physical attacks. These sequences of data begin 400 seconds prior to when an attack begins. For attacks with windows smaller than 400 seconds between each other we use the largest of either a 100 or 25 second window. We include the attack-free data at the start of the sequence to ensure that the LSTM internal state has stabilised from its initial values.  

During the definition of clean data sequences we tried different values of the time parameters just mentioned, and observed little to no change in the results.

We run these sequences of data through our model and tune the thresholds using the same procedure as in section \ref{sec:IDS_Results}. We generate two different sets of thresholds for the attacker to overcome. The first is for the case where the attacker targets only continuous data, and the IDS monitors only the continuous sensors. The second is for the more challenging case where the attacker targets a mixture of continuous and discrete data, and the IDS monitors everything. 

In terms of recall, which is a good measure of how difficult a target the IDS will be for an attacker, we obtain a score of 0.821 when solely monitoring the continuous data. When monitoring both types of features we achieve a recall of 0.834. This places us in the same region as the strongest literature results for dedicated defence papers (0.821/0.834 (ours) vs 0.827 \cite{kravchik2019efficient}). In terms of $F_1$ scores, monitoring only the continuous sensors achieves an $F_1$ of 0.872 while monitoring both continuous and discrete data obtains an $F_1$ of 0.886. We should emphasise that these $F_1$ scores are \textit{not} comparable to the results in Table \ref{tab:F1_Scores}, as we are computing the scores by removing sections of data in the procedure as described earlier. What they do show (particularly in terms of recall) is that the defender IDS is now a more significant hurdle for the attacker. 

\begin{figure}
\centering
\includegraphics[width=0.8\linewidth]{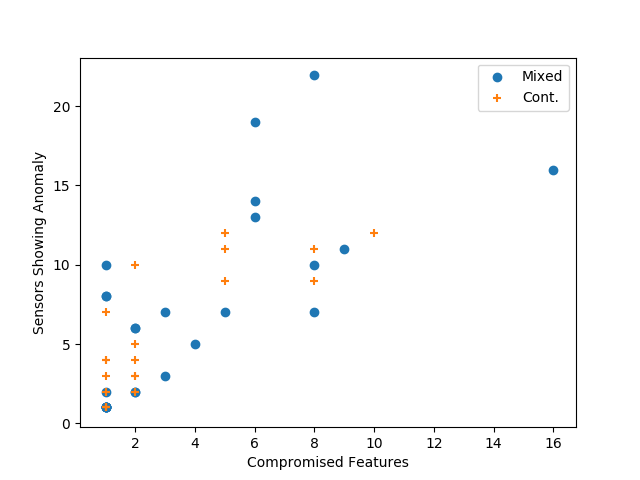}
\caption{Number of features which caused alerts against the level of compromise needed to hide the attack for continuous-only and mixed data.}
\label{fig:L0_Results}
\end{figure}

\section{Adversarial Results}

The full set of results is presented in Appendix Table \ref{table:main_results} in which we show the minimum compromise needed to hide every cyber-physical attack. Although there are 36 total attacks in the SWaT dataset, two separate pairs occur back to back. Here we consider them forming single longer attacks as it would be unrealistic for an attacker to begin optimising from scratch in the middle of a detected attack. In the continuous data regime 24 out of 34 cyber-physical attacks are successfully detected by the IDS. From the detected 24 cyber-physical attacks we successfully hide 23 of them with our adversarial attack. For the mixed data regimes, 29 out of 34 attacks are detected by the IDS and of those our adversarial attacks hide 27.

The adversarial attacks which are unsuccessful, one in the continuous data regime and two cases when considering mixed data, fail due to unknown sensor noise present in the defender data $D_d$ as described in section \ref{sec:sensor_noise}.

Regarding the amount of compromise required, on average 2.87 compromised features are required to hide a cyber-physical attack out of 12 monitored features when dealing with purely continuous data. In the mixed continuous/discrete regime we require 3.74 features to be compromised out of 26 monitored features. In general, the more features over which an attack is detected, the more features need compromising: for the continuous data scenario scenario, on average attacks which are detected on 6 or more features require 5.63 features to be compromised, while attacks which are detected on less than 6 require 1.4 compromised features. From Figure \ref{fig:L0_Results} we see the number of features an attack is detected on against the number of features requiring compromise.

\begin{figure*}
\centering
\subfloat[Original readings for LIT301.]{\includegraphics[width=.3\linewidth]{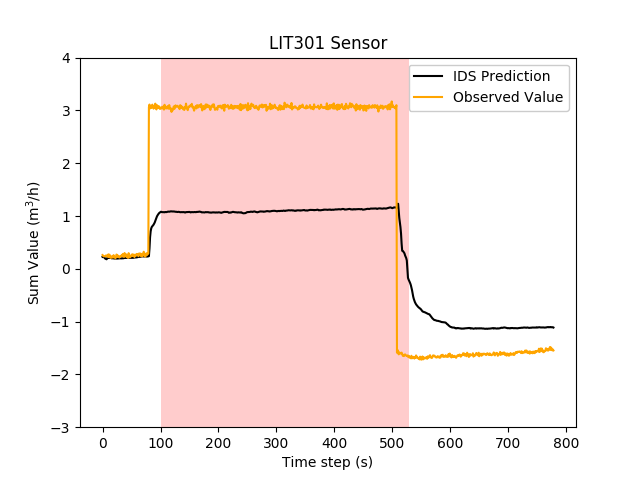}\label{fig:sfig1}} \qquad
\subfloat[Original readings for FIT201.]{\includegraphics[width=.3\linewidth]{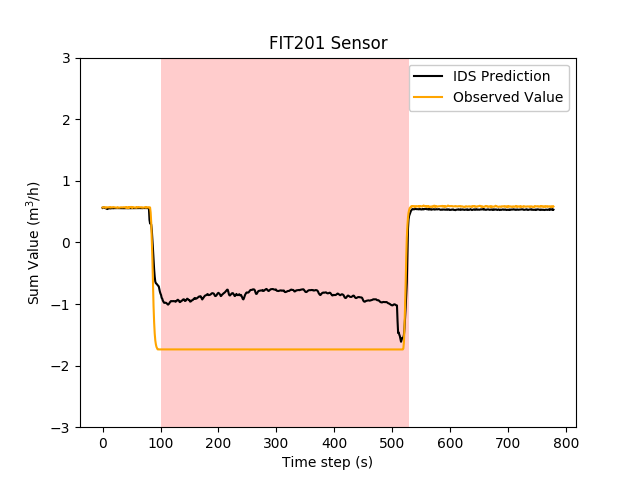}\label{fig:sfig2}} \\
\subfloat[The residuals accumulate to trigger a alert on the LIT301 sensor.]{\includegraphics[width=.3\linewidth]{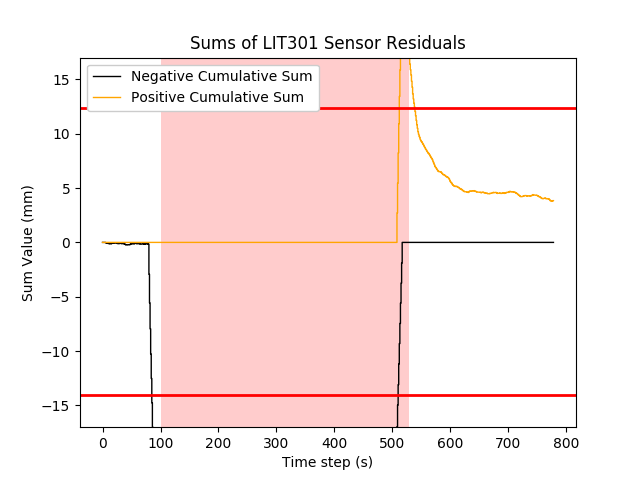}\label{fig:sfig3}} \qquad
\subfloat[The residuals accumulate to trigger an alert on the FIT201 sensor.]{\includegraphics[width=.3\linewidth]{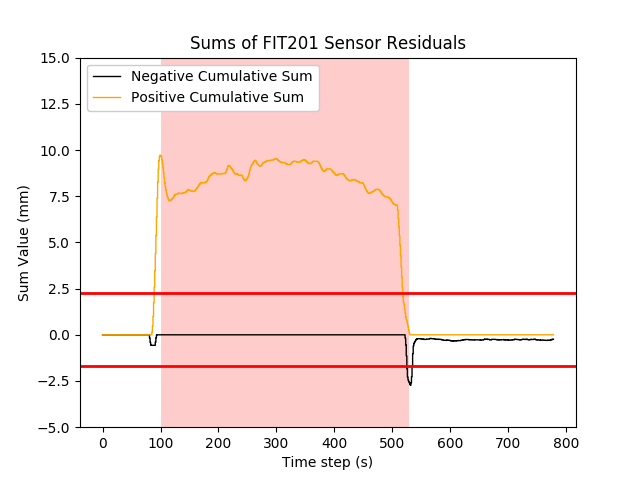}\label{fig:sfig4}} \\
\subfloat[LIT301 readings when an adversary optimises to reduce the residuals to below detection across all sensors.]{\includegraphics[width=.3\linewidth]{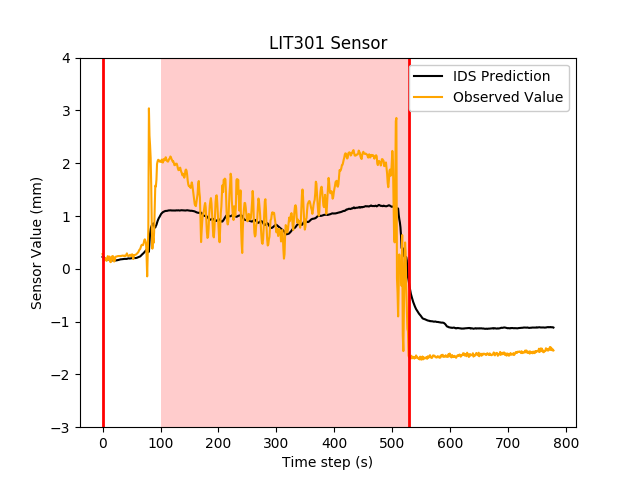}\label{fig:LIT301_adv_sensor}} \qquad
\subfloat[FIT201 readings when an adversary optimises to reduce the residuals to below detection across all sensors.]{\includegraphics[width=.3\linewidth]{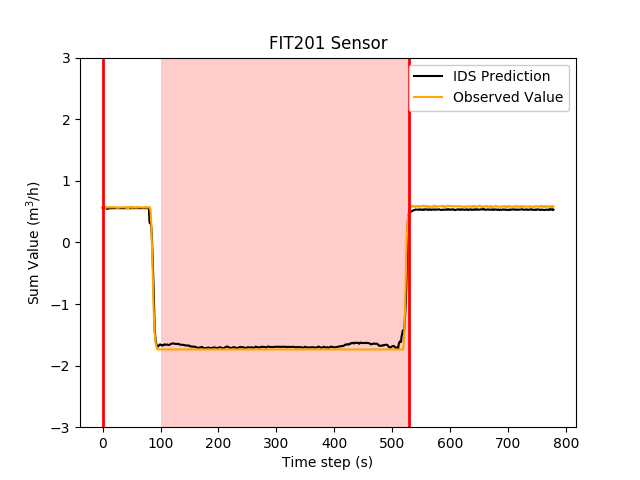}\label{fig:FIT201_adv_sensor}} \\
\subfloat[The LIT301 residual sums resulting from the adversarial attack.]{\includegraphics[width=.3\linewidth]{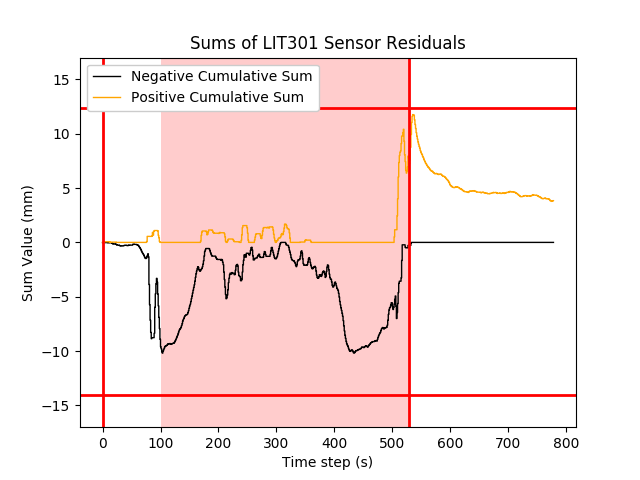}\label{fig:LIT301_adv_res}} \qquad
\subfloat[The FIT201 residual sums resulting from the adversarial attack.]{\includegraphics[width=.3\linewidth]{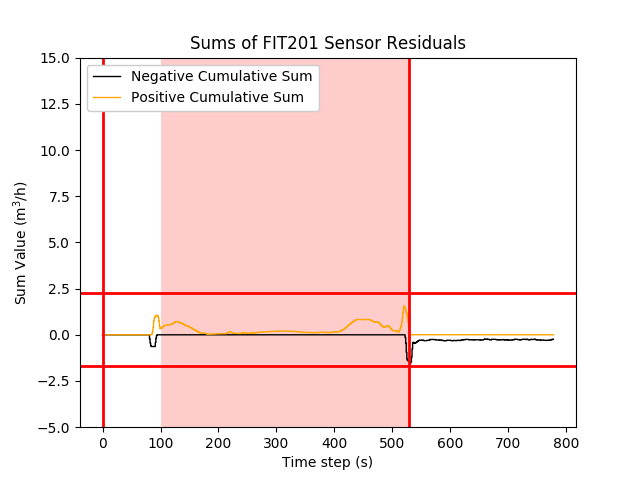}\label{fig:FIT201_adv_res}}

\label{fig:Attack_6_Normal}
\caption{Results for the LIT301 (left plots) and FIT201 sensors (right plots) for attack number 6. Red vertical lines indicate when the attacker began and ended their optimisation and the shaded area indicates when the cyber-physical attack is in progress. Horizontal red lines show detection thresholds.}
\end{figure*}

To gain a better understanding of how the attack modifies the data we show a portion of attack number 6 which aims to cause an underflow in a tank. The raw data is shown in in Figures \ref{fig:sfig1} - \ref{fig:FIT201_adv_res} and here we assume that the attacker can control a single sensor (LIT301). Figures \ref{fig:sfig1} and \ref{fig:sfig2} show how the readings for sensors LIT301 (the compromised sensor) and FIT201 (one of the 4 sensors that triggered an anomaly) differ from the predicted values generating residuals in Figures \ref{fig:sfig3}-\ref{fig:sfig4}. The attacker compromises the LIT301 sensor and optimises to keep the residuals hidden across all of the sensors which triggered an anomaly (LIT301, LIT101, FIT201, FIT504). In Figure \ref{fig:LIT301_adv_sensor} we see how the adversarial data on the LIT301 sensor needs be perturbed to achieve this. Then, in Figure \ref{fig:FIT201_adv_sensor} we see the effect this has had on the predicted values for FIT201 and the residuals in Figures \ref{fig:LIT301_adv_res}-\ref{fig:FIT201_adv_res} have been reduced to as to not trigger an alert.

The most similar work to ours is \cite{erba2019real} which also constructs adversarial examples for SWaT, however direct comparison between results is challenging as 1) their attacker model operates on different perturbation constraints 2) the intrusion detection system is non-autoregressive which is a key focus of this paper and 3) the attacker objective also differed, we seek to reduce detectability to 0 while in \cite{erba2019real} the detection magnitude is reduced as much as possible (but not to 0) within their attacker budget dependent on their perturbation constraints.

\subsection{Replay Attack Comparison}
 
The replay attack baseline performs significantly worse than our approach. The average level of compromise needed when considering mixed datatypes is 10.26 sensors (compared to 3.74 sensors for our attack strategy) and fails to hide 6 attacks due to unknown sensor noise, while our algorithm fails on 2 attacks. When only being able to compromise continuous data a replay attack strategy is ineffective functioning only on 2 of the examined attacks. This is because without a degree of optimisation which considers the discrete features, which remain fixed, the IDS detects the discrepancies between the discrete and continuous features even if the continuous features come from attack free data.
 
\subsection{Transferability of Adversarial Examples}

With the adversarial examples showing sensitivity to sensor noise, we examine what would occur if the attacker had black box knowledge of the IDS. Specifically, the attacker knows the training/validation data, the model architecture, and the algorithm used to generate the thresholds. To examine this we train a new detection model and evaluate the attacks that have been generated for the original IDS on the new model. 

When examining only continuous data 18 of the 23 successful attacks transferred. There is a correlation between the attacks that used the $L_0$ Prediction attack strategy and its transferability. Only 2 attacks which utilised the $L_0$ Prediction attack strategy in some form failed to transfer. This can be explained as by purely using our $L_0$ Optimisation attack, the goal is only to remain under a fraction $\tau$ of the detection threshold. This places the residuals only slightly under detectability. However, initialising the optimisation via the $L_0$ Prediction strategy and then, if necessary, optimising further places the results a larger margin under the detection threshold. An example of this combined strategy on the residuals is shown in Figures \ref{fig:LIT301_adv_sensor} - \ref{fig:FIT201_adv_res}.

When both discrete and continuous data is considered 15 out of the 27 successful adversarial attacks transferred. This is a deterioration in performance compared to the purely continuous data regime, and reflects the more challenging underlying optimisation task that is required.

\section{Conclusion}

We have presented a method for generating adversarial attacks on time-series IDS. Although they can be fooled a higher degree of perturbation is required along with stronger optimisation strategies in comparison to the image domain.

A key difficulty with generating the adversarial attacks is due to the attacker perturbations influencing the optimisation target at every iteration step requiring a search for the best step size to be conducted. This problem becomes more pronounced when several features are being optimised as they add additional shifts to the optimisation target.

\ifCLASSOPTIONcompsoc
  \section*{Acknowledgments}
\else
  \section*{Acknowledgment}
\fi

The authors would like to thank NVIDIA for the donation of a GPU in support of this work. This work is funded by a joint scholarship between EPRSC and Airbus.

\bibliographystyle{IEEEtran}
\bibliography{Ref}
\section*{Appendix}

\label{sec:Naming_Conventions}
Occasionally a specific sensor or actuator is referred to in this work. To assist in understanding what type of senor or actuator is being referred to their roles are as follows: 
\begin{itemize}
    \item LIT: Water Level Indicator Transmitter. 
    \item FIT: Water Flow Indicator Transmitter. 
    \item MV: Mechanical Valve actuator. 
    \item P: Pump actuator. 
    \item AIT: Analyser Indicator Transmitter. Measures one of conductivity, pH, or oxidation reduction potential.  
    \item DPIT: Differential Pressure Indicator Transmitter.
\end{itemize}

Additionally a numerical suffix is usually appended to identify which of the many different pumps, valves, or transmitters is being referred to. The first digit identifies in which sub-process the component is located and the second two digits are its numerical indicator. Thus, LIT301 is a sensor measuring the water level in sub-process 3 and is the first numerically listed.

\begin{table*}[!b]
    We state for every cyber-physical attack in the SWaT testbed 1) The number of features that detect an anomaly. 2) The number of features that require compromise to hide the cyber-physical attack completely. Finally, 3) which of the attack strategies is used in the case of purely continuous (Cont.) data being examined (the mixed data domain always uses the same strategy). Attack pairs 8 / 9  and 22 / 23 occur back to back in the attack dataset and we consider the them together forming as a single longer attack. For the mixed continuous and discrete data regime we always use the same attack method as described in \ref{sec:Attack_Strategy}. ``F" in the ``Compromised Features to Hide Attack" column indicated the attack failed due to sensor noise. In total there are 12 monitored sensors when considering purely continuous data and 26 monitored sensors when using both continuous and discrete data.
    \newline
    
	\caption{Complete Table of Results.}
    \label{table:main_results}
    \centering
		\begin{tabular}{ c c p{1.2cm} p{1.2cm} p{1.5cm} p{1.5cm} c}
			\toprule
			
			Attack & Attack Description & \multicolumn{2}{c}{\begin{tabular}{@{}c@{}}Features with \\ Detection\end{tabular}} & \multicolumn{2}{c}{\begin{tabular}{@{}c@{}}Compromised Features \\ to Hide Attack\end{tabular}} &
			\begin{tabular}{@{}c@{}}Adversarial Method \\  Used for Cont.\end{tabular} \\
			
			      & &\centering Cont. & \centering Mixed  & \centering Cont. & \centering Mixed & \\

			\hline
			\addlinespace[0.15cm]
			1 & \begin{tabular}{@{}c@{}}Open MV101 to \\  overflow tank\end{tabular} &\centering 2 & \centering 2 & \centering 1 & \centering 2 & $L_0$ Prediction  \\
			\addlinespace[0.15cm]
			2 & \begin{tabular}{@{}c@{}}Turn on P102 to \\  bust pipes\end{tabular}
			 &\centering 1 & \centering 3 & \centering 1 & \centering 3 & $L_0$ Prediction\\
			\addlinespace[0.15cm]
			3 & \begin{tabular}{@{}c@{}}Increase LIT101 to \\ underflow tank\end{tabular}
			&\centering 1 & \centering 1 & \centering 1 & \centering 1 & $L_0$ Prediction\\
			\addlinespace[0.15cm]
			4 & \begin{tabular}{@{}c@{}}Open MV504 to halt\\ reverse osmosis shutdown\end{tabular}
			&\centering 0 & \centering 0 & \centering NA & \centering NA & NA\\
			\addlinespace[0.15cm]
			5 & \begin{tabular}{@{}c@{}} Tamper AIT202 to\\ reduce water quality\end{tabular} &\centering 0 & \centering 1 & \centering NA & \centering 1 & NA\\
			\addlinespace[0.15cm]
			6 & \begin{tabular}{@{}c@{}}Increase LIT301 to \\ underflow tank\end{tabular} &\centering 4 & \centering 6 & \centering 1 & \centering 2 & \begin{tabular}{@{}c@{}}$L_0$ Prediction\\and $L_0$ Optimisation\end{tabular}\\
			\addlinespace[0.15cm]
			7 & \begin{tabular}{@{}c@{}}Increase value of DPIT to\\stop system operation\end{tabular} &\centering 9 & \centering 11 & \centering 9 & \centering 9 & \begin{tabular}{@{}c@{}}$L_0$ Prediction\\and $L_0$ Optimisation\end{tabular}\\
			\addlinespace[0.15cm]
			8 + 9 & \begin{tabular}{@{}c@{}}Reduce value of FIT401 to \\disrupt system operation\end{tabular} &\centering 11 & \centering 22 & \centering 8 & \centering 8 & \begin{tabular}{@{}c@{}}$L_0$ Prediction\\and $L_0$ Optimisation\end{tabular}\\
			\addlinespace[0.15cm]
			10 & \begin{tabular}{@{}c@{}}Close MV304 to \\ halt stage 3\end{tabular}  &\centering 0 & \centering 0 & \centering NA & \centering NA & NA\\
			\addlinespace[0.15cm]
			11 & \begin{tabular}{@{}c@{}}Do not open MV303 to \\ halt stage 3\end{tabular} &\centering 0 & \centering 0 & \centering NA & \centering NA & NA\\
			\addlinespace[0.15cm]
			12 & \begin{tabular}{@{}c@{}}Decrease LIT301 to \\ overflow tank\end{tabular} &\centering 2 & \centering 2 & \centering 2 & \centering 2 & $L_0$ Prediction\\
			\addlinespace[0.15cm]
			13 & \begin{tabular}{@{}c@{}}Do not open MV303 to \\ halt stage 3\end{tabular} &\centering 3 & \centering 7 & \centering 2 & \centering 8 & $L_0$ Optimisation\\
			\addlinespace[0.15cm]
			14 & \begin{tabular}{@{}c@{}}Increase AIT504 to \\ cause drain\end{tabular} &\centering 0 & \centering 1 & \centering NA & \centering 1 & NA\\
			\addlinespace[0.15cm]
			15 & \begin{tabular}{@{}c@{}}Increase AIT504 to \\ cause drain\end{tabular} &\centering 0 & \centering 0  & \centering NA & \centering NA & NA\\
			\addlinespace[0.15cm]
			16 & \begin{tabular}{@{}c@{}}Keep MV101 on. Decrease\\ LIT101 to overflow tank\end{tabular} &\centering 1 & \centering 1 & \centering 1 & \centering 1 & $L_0$ Prediction\\
			\addlinespace[0.15cm]
			17 & \begin{tabular}{@{}c@{}}Multi-point attack to \\ damage reverse osmosis\end{tabular}&\centering 11 & \centering 13 & \centering 5 & \centering 6 & \begin{tabular}{@{}c@{}}$L_0$ Prediction\\and $L_0$ Optimisation\end{tabular}\\
			\addlinespace[0.15cm]
			18 & \begin{tabular}{@{}c@{}}Multi-point attack to \\ freeze system\end{tabular} &\centering 0 & \centering 10 & \centering NA & \centering 8 & NA\\
			\addlinespace[0.15cm]
			19 & \begin{tabular}{@{}c@{}}Turn off P203 and P205 \\ change water quality\end{tabular} &\centering 0 & \centering 1 & \centering NA & \centering 1 & NA\\
			\addlinespace[0.15cm]
			20 & \begin{tabular}{@{}c@{}}Increase LIT401 and keep\\P402 on to underflow tank\end{tabular} &\centering 3 & \centering 1 & \centering 1 & \centering 1 & $L_0$ Prediction\\
			\addlinespace[0.15cm]

			\end{tabular}
	\end{table*}

\begin{table*}[!t]
	\caption{Continuation of results table.}
    \centering
		\begin{tabular}{ c c p{1.2cm} p{1.2cm} p{1.5cm} p{1.5cm} c}
			\toprule
			
			Attack & Attack Description & \multicolumn{2}{c}{\begin{tabular}{@{}c@{}}Features with \\ Detection\end{tabular}} & \multicolumn{2}{c}{\begin{tabular}{@{}c@{}}Compromised Features \\ to Hide Attack\end{tabular}} &
			\begin{tabular}{@{}c@{}}Adversarial Method \\  Used for Cont.\end{tabular} \\
			
			      & &\centering Cont. & \centering Mixed  & \centering Cont. & \centering Mixed & \\

			\hline
			\addlinespace[0.15cm]			
			21 & \begin{tabular}{@{}c@{}}Multi-point attack to \\ damage two tanks\end{tabular}&\centering 4 & \centering 7 & \centering 2 & \centering F & $L_0$ Prediction\\
	        \addlinespace[0.15cm]
			22 + 23  & \begin{tabular}{@{}c@{}}22: Tank overflow \\ 23: Stop tank inflow\end{tabular} &\centering 12 & \centering 16 & \centering 10 & \centering 16 & \begin{tabular}{@{}c@{}}$L_0$ Prediction\\and $L_0$ Optimisation\end{tabular}\\
			\addlinespace[0.15cm]
			24 & \begin{tabular}{@{}c@{}}Turn on P201, P203, and\\P205 to waste chemicals\end{tabular} &\centering 0 & \centering 0 & \centering NA & \centering NA & NA\\
			\addlinespace[0.15cm]
			25 & \begin{tabular}{@{}c@{}}Turn on P101 and MV101 \\ to underflow/overflow two tanks\end{tabular} &\centering 3 & \centering 7 & \centering 2 & \centering 3 & $L_0$ Prediction\\
			\addlinespace[0.15cm]
			26 & \begin{tabular}{@{}c@{}}Reduce LIT401 \\to overflow tank\end{tabular} &\centering 5 & \centering 8 & \centering F & \centering 1 & NA \\
			\addlinespace[0.15cm]
			27 & \begin{tabular}{@{}c@{}}Increase LIT301\\ to underflow tank\end{tabular} &\centering 4 & \centering 6 & \centering 1 & \centering 2 & $L_0$ Optimisation\\
			\addlinespace[0.15cm]
			28 & \begin{tabular}{@{}c@{}}Increase LIT101\\ to underflow tank\end{tabular}&\centering 3 & \centering 7 & \centering 2 & \centering 5 & \begin{tabular}{@{}c@{}}$L_0$ Prediction\\and $L_0$ Optimisation\end{tabular}\\
			\addlinespace[0.15cm]
			29 & \begin{tabular}{@{}c@{}}Turn off P101\\ to stop outflow\end{tabular} &\centering 0 & \centering 1 & \centering NA & \centering 1 & NA\\
			\addlinespace[0.15cm]
			30 &\begin{tabular}{@{}c@{}}Turn off P101 and P102\\ to stop outflow\end{tabular} &\centering 1 & \centering 5 & \centering 1 & \centering 4 & $L_0$ Prediction\\
			\addlinespace[0.15cm]
			31 & \begin{tabular}{@{}c@{}}Reduce LIT101\\ to overflow tank\end{tabular} &\centering 5 & \centering 8 & \centering 2 & \centering 1 & $L_0$ Optimisation\\
			\addlinespace[0.15cm]
			32 & \begin{tabular}{@{}c@{}}Close P501 and vary FIT502\\ to reduce output\end{tabular}&\centering 12 & \centering 19 & \centering 5 & \centering 6 & \begin{tabular}{@{}c@{}}$L_0$ Prediction\\and $L_0$ Optimisation\end{tabular}\\
			\addlinespace[0.15cm]
			33 & \begin{tabular}{@{}c@{}}Manipulate AIT502\\ to send water to drain\end{tabular} &\centering 2 & \centering 2 & \centering 1 & \centering 2 & $L_0$ Optimisation\\
			\addlinespace[0.15cm]
			34 & \begin{tabular}{@{}c@{}}FIT401 and AIT502 manipulation\\to disrupt UV and reverse osmosis\end{tabular} &\centering 10 & \centering 8 & \centering 2 & \centering F & \begin{tabular}{@{}c@{}}$L_0$ Prediction\\and $L_0$ Optimisation\end{tabular}\\
			\addlinespace[0.15cm]
			35 &  \begin{tabular}{@{}c@{}}Decrease FIT401 to disrupt\\UV and reverse osmosis \end{tabular} &\centering 9 & \centering 14 & \centering 5 & \centering 5 & \begin{tabular}{@{}c@{}}$L_0$ Prediction\\and $L_0$ Optimisation\end{tabular}\\
			\addlinespace[0.15cm]
			36 & \begin{tabular}{@{}c@{}}Decrease LIT301 to\\ overflow tank\end{tabular} &\centering 7 & \centering 10 & \centering 1 & \centering 1 & $L_0$ Prediction\\
		    \bottomrule
		    \addlinespace[1cm]
			\end{tabular}
	\end{table*}

\end{document}